\documentclass[aps,prl,superscriptaddress,twocolumn]{revtex4}
\usepackage{amsmath}
\usepackage{graphicx}
\usepackage{color}

\begin{document}

\newcommand{\beq}{\begin{equation}}
\newcommand{\eeq}{\end{equation}}
\newcommand{\beqa}{\begin{eqnarray}}
\newcommand{\eeqa}{\end{eqnarray}}
\newcommand{\ben}{\begin{enumerate}}
\newcommand{\een}{\end{enumerate}}
\newcommand{\hs}{\hspace{0.5cm}}
\newcommand{\vs}{\vspace{0.5cm}}
\newcommand{\note}[1]{{\color{red} \bf [#1]}}

\title{Entanglement at a Two-Dimensional Quantum Critical Point: \\
a Numerical Linked Cluster Expansion Study}

\author{Ann B. Kallin}
\affiliation{Department of Physics and Astronomy, University of Waterloo, Ontario, N2L 3G1, Canada}

\author{Katharine Hyatt}
\affiliation{Department of Physics and Astronomy, University of Waterloo, Ontario, N2L 3G1, Canada}
\affiliation{Department of Physics, University of California, Santa Barbara, CA, 93106-9530}

\author{Rajiv R. P. Singh}
\affiliation{Physics Department, University of California, Davis, CA, 95616}

\author{Roger G. Melko}
\affiliation{Department of Physics and Astronomy, University of Waterloo, Ontario, N2L 3G1, Canada}
\affiliation{Perimeter Institute for Theoretical Physics, Waterloo, Ontario N2L 2Y5, Canada}

\date{\today}

\begin{abstract}
We develop a method to calculate the bipartite entanglement entropy of 
quantum models, in the thermodynamic limit, using a Numerical Linked 
Cluster Expansion (NLCE) involving only rectangular clusters.
It is based on exact diagonalization of all $n\times m$ rectangular clusters 
at the interface between entangled subsystems $A$ and $B$. 
We use it to obtain the Renyi 
entanglement entropy of the two-dimensional transverse field Ising 
model, for arbitrary real Renyi index $\alpha$.  Extrapolating
these results as a function of the order of the calculation, we obtain
universal pieces of the entanglement entropy associated with lines
and corners at the quantum critical point. 
They show NLCE to be one of the few methods capable of accurately calculating
universal properties of arbitrary Renyi entropies at higher dimensional critical points.
\end{abstract}

\maketitle

{\em Introduction --}
Quantum critical points (QCPs) \cite{Sachdev11} offer some of the most non-classical, or highly-entangled, states in 
condensed matter physics.  Although this high degree of entanglement can be a challenge for efforts
to construct general numerical methods to study QCPs \cite{MERA1,MERA_QCP}, it can also be viewed as a resource to detect
and classify them.  In particular, it is believed that the sub-leading 
scaling terms of the Renyi entanglement entropies contain universal coefficients \cite{logcorner,Max}.  Thus, these universal
terms can be studied in quantum many-body models using numerical techniques, such as quantum
Monte Carlo (QMC) or exact diagonalization, and compared to quantum field theories as a way of determining
the universality class of a QCP.  In addition to conventional universality classes, this procedure
has the potential to identify unconventional (non-Landau) QCPs, which are predicted to be even more
highly entangled than their conventional counterparts \cite{Swingle_DQCP}.  Thus, it is important to have unbiased numerical
methods which can calculate these universal numbers for a quantitative comparison to field theories,
models, and perhaps some day, experimental studies \cite{Cardy_PRL,Abanin}.

\begin{figure}[ht]
\centering
\includegraphics*[width=2.5in]{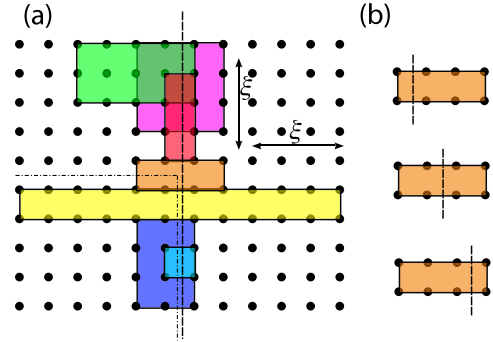}
\caption[]{
(a) A sample of cluster shapes and sizes, as used in the rectangular NLCE.  
(b) To calculate the Renyi entropies of a given cluster, one considers all
divisions into subregions $A$ and $B$ defined by 
translations perpendicular to the line (illustrated) or corner.
\label{Rectangles}       
}
\end{figure}

In this paper, 
we study the entanglement of a two-dimensional (2D) transverse-field Ising model at its quantum critical point.  
We use the Numerical Linked Cluster Expansion (NLCE) \cite{NLC1,NLC2,NLC3} to calculate the 
Renyi entanglement entropy \cite{A_renyi} for arbitrary index $\alpha$.
The NLCE is based on an exact diagonalization of finite-size clusters 
up to some ``order'' $\mathcal{O}$ corresponding to the number of sites in each cluster.
By introducing an innovation that allows us to only consider rectangular clusters, we are able to perform 
NLCE up to unprecedented orders.  Through direct calculation of universal properties associated with
entanglement across corners, we demonstrate that the accuracy of NLCE rivals (or even bests) other numerical methods 
including quantum Monte Carlo, tensor tree networks, and series expansions.  
We show that the universal term in the entanglement entropy associated with a corner
is distinct from the value calculated in a non-interacting field theory \cite{logcorner}.  
We also use the unique ability of NLCE to calculate Renyi entropies with non-integer $\alpha$ values
to search for a striking change of sign in the universal coefficient of line entropy, predicted by an interacting 
field theory near $\alpha = 1$ \cite{Max}.  We conclude that no sign change takes place, suggesting that either
this phenomena occurs at exceptionally long length scales, or, more likely,
low-order perturbative expansions in the field theory are inadequate for the Ising universality class in $(2+1)D$.

Numerical linked cluster expansion (NLCE) \cite{NLC1,NLC2,NLC3} is based on expressing an extensive property $P$ of a 
lattice model, per site, as a sum over contributions from all distinct clusters $c$ that can be 
embedded in the lattice:
\begin{equation}
P/N = \sum_c L(c) \times W(c) \label{eqP}
\end{equation} 
where $L(c)$ is the number of embeddings of the cluster per lattice site, 
and $W(c)$ is the ``weight'' of the cluster for the
property $P$, defined according to the inclusion-exclusion principle,
\begin{equation}
W(c) = P(c) - \sum_{s \in c} W(s). \label{eqW}
\end{equation}
Here, $P(c)$ is the property calculated for a finite cluster $c$, which contains sub-clusters $s$.  
In the NLCE method the property is calculated using an exact diagonalization for each cluster, 
and the calculations are carried out up to some order $\mathcal{O}$, the number of sites in the cluster.

In Eq.~(\ref{eqW}), the weight of every cluster captures contributions to $P$ from correlations 
contained within the size of the cluster.  When the correlation length grows larger than the
largest cluster size considered, for example at a quantum critical point, results from NLCE 
require an extrapolation in the size of the clusters.
Thus, it is important to consider as large clusters as possible. 

Typical NLCE approaches involve constructing clusters that are site- or bond-based. 
These can involve
the computationally expensive task of generating all clusters $c$ and subclusters $s$ up to a
given order $\mathcal{O}$ --
a procedure that is related
to an NP-complete graph embedding problem \cite{Cook}.  When calculating groundstate properties, e.g.~using Lanczos
diagonalization, this graph embedding problem is the computational bottleneck, restricting
the order of conventional expansions to $\sim16$, even with clever use of point-group symmetries 
and topology to reduce graph counting \cite{NLC4}.  
In this paper, we take advantage of
the ability of the NLCE procedure, Eq.~(\ref{eqP}), to converge with alternate definitions of
cluster geometries, as long as each cluster $c$ can be self-consistently decomposed into 
subclusters $s$ also defined in this alternate way, according to Eq.~(\ref{eqW}).
Then, a particularly convenient choice for cluster geometry on the square lattice is
$m \times n$ site {\it rectangles} (Fig.~\ref{Rectangles}), 
since any general bond- or site-based cluster can be assigned to a unique rectangle.
This significantly reduces 
the number of clusters required to self-consistently define the sum,
and makes the counting of graphs and subgraphs trivial \cite{Enting,Dusuel}.  
The computational bottleneck then becomes the exponential Hilbert space 
of each rectangular cluster, which is stored in memory during the Lanczos diagonalization.
This allows us to push the NLCE to significantly higher orders -- up to ${\mathcal O}=26$ with moderate effort on simple desktop workstations \footnote{Slightly larger orders are clearly possible with a moderate increase in effort.}.
We stress that the NLCE is not a calculation for a finite size system, rather a systematic approximation for 
the thermodynamic limit, where the rectangles provide a way to sum up contributions 
from different length scales corresponding to the order of the cluster.
NLCE can systematically encapsulate significantly larger-range correlations than
conventional finite-size studies of toroidal clusters (as is often done in Lanczos diagonalization)
which are crucial to determine the singular behavior at a QCP \footnote{
We have benchmarked our NLCE procedure in the case of $1D$, 
which is however quite well understood and hence will not be discussed further.}.

We use this method to study the scaling of the generalized Renyi
entropies at the quantum critical point of the 2D transverse field Ising model (TFIM),
\begin{equation}
H = -J\sum_{\langle i,j \rangle} \sigma^z_i \sigma^z_j - h \sum_{i} \sigma^x_i,
\end{equation}
where ${\overrightarrow{\sigma}}_i$ is a Pauli spin operator, so that $\sigma^z$ has eigenvalues $\pm 1$.  
In this equation, the first sum is over lattice {\em bonds}, while the second sum is over lattice {\em sites}.
To calculate the Renyi entanglement entropies, $S_{\alpha} = \ln({\rm Tr} \rho_A^{\alpha})/(1-\alpha)$ 
one must position each cluster in relation to the boundary between subregions $A$ and $B$,
and calculate the reduced density matrix for the subsystem $A$.  Translational symmetry along
the boundary simply gives an overall factor of length $L$, and automatically produces the 
entropy per unit length associated with a line, when only translationally distinct clusters are included. To take care
of translations perpendicular to the line, or translations with respect to a corner, it is
equivalent and more convenient to consider a given $n\times m$ cluster only once, but
allow all possible linear (in the case of a line, illustrated in Fig.~\ref{Rectangles}(b)) or 
quadrant-based divisions of that cluster.

A key advantage of the NLCE method over quantum Monte Carlo (QMC) \cite{XXZ}
and series expansions \cite{TFIM_series} is that
it can be used to calculate Renyi entropies for any $\alpha$ value including $\alpha \le 1$. Lanczos diagonalization
provides us with the exact ground state wavefunction of each cluster, allowing an explicit calculation of
the reduced density matrix, from which Renyi entropies with arbitrary $\alpha$ can be obtained.
Another distinct advantage of the NLCE method over QMC is that
one can analytically separate the Renyi entropies associated with lines and corners.
When the subsystem $A$ is a half-plane, one obtains only the entropy associated with the line.
When the subsystem $A$ is a quadrant, it contains both line and corner contributions.
A suitable choice of subdivision of the system into half-planes and corners is sufficient to isolate the corner
contribution from every graph, thus leading to a separate calculation for line and corner entropies for the TFIM, that we call
$s_{\alpha}=S_{\alpha}/L$ and $c_{\alpha}$ below. This, in turn, allows a more accurate determination of the singularities
associated with each term than possible in QMC, where e.g.~the dominant ``area law'' can easily overwhelm 
sub-leading terms such as corner contributions.

\begin{figure}[t]
\includegraphics*[width=3.0in]{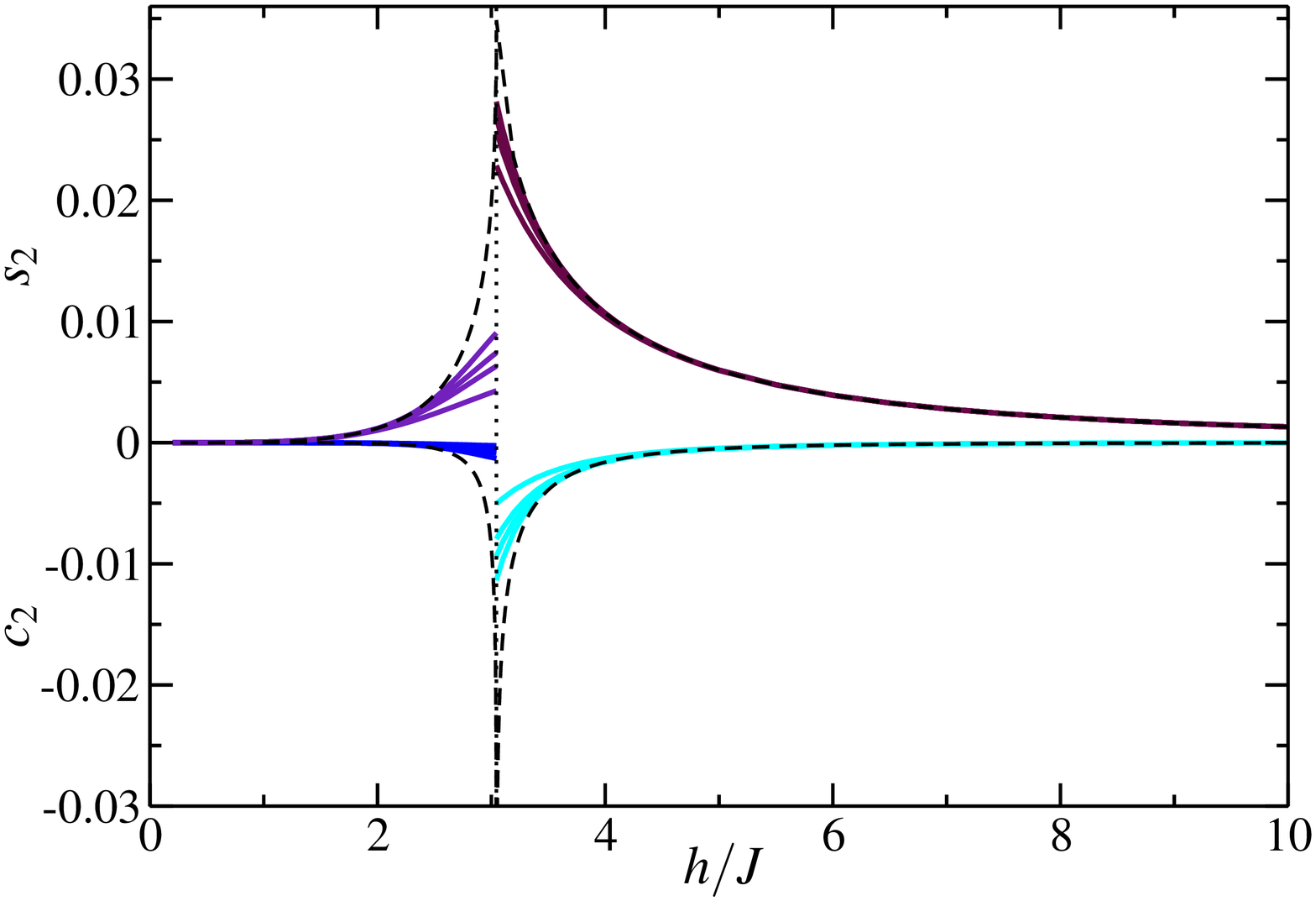}
\caption[]{The corner term $c_2$ and line term $s_2$ as a function of $h/J$ for the second Renyi entanglement entropy using both high field and low field NLCE of orders 8, 12, 16, and 24.  The dotted line denotes $h_c/J=3.044$ and the dashed line shows the series expansion results \cite{TFIM_series}.
\label{cornerline}      
}
\end{figure}

In Fig.~\ref{cornerline} we show results for line and corner entropies for the TFIM as a function of $h/J$
at different orders, where $h_c/J=3.044$ is the quantum critical point.
Previously, the series expansion method was used to calculate $S_2$ \cite{TFIM_series}, but
not for example $S_1$, which is also inaccessible to QMC calculations due to their reliance on the replica trick. Note that,
in order to get convergence for $h<h_c$, one needs to add a static ordered moment for
the sites outside the cluster. These moments apply a boundary field on the sites in the cluster,
via the exchange $J$, and ensure that fluctuations remain bounded. 
This is analogous to low field series expansions, and we call it {\em low field} NLCE. For $h\ge h_c$ no such
boundary field is needed and we call it {\em high field} NLCE. 

\begin{figure}[t]
\includegraphics*[width=3.0in]{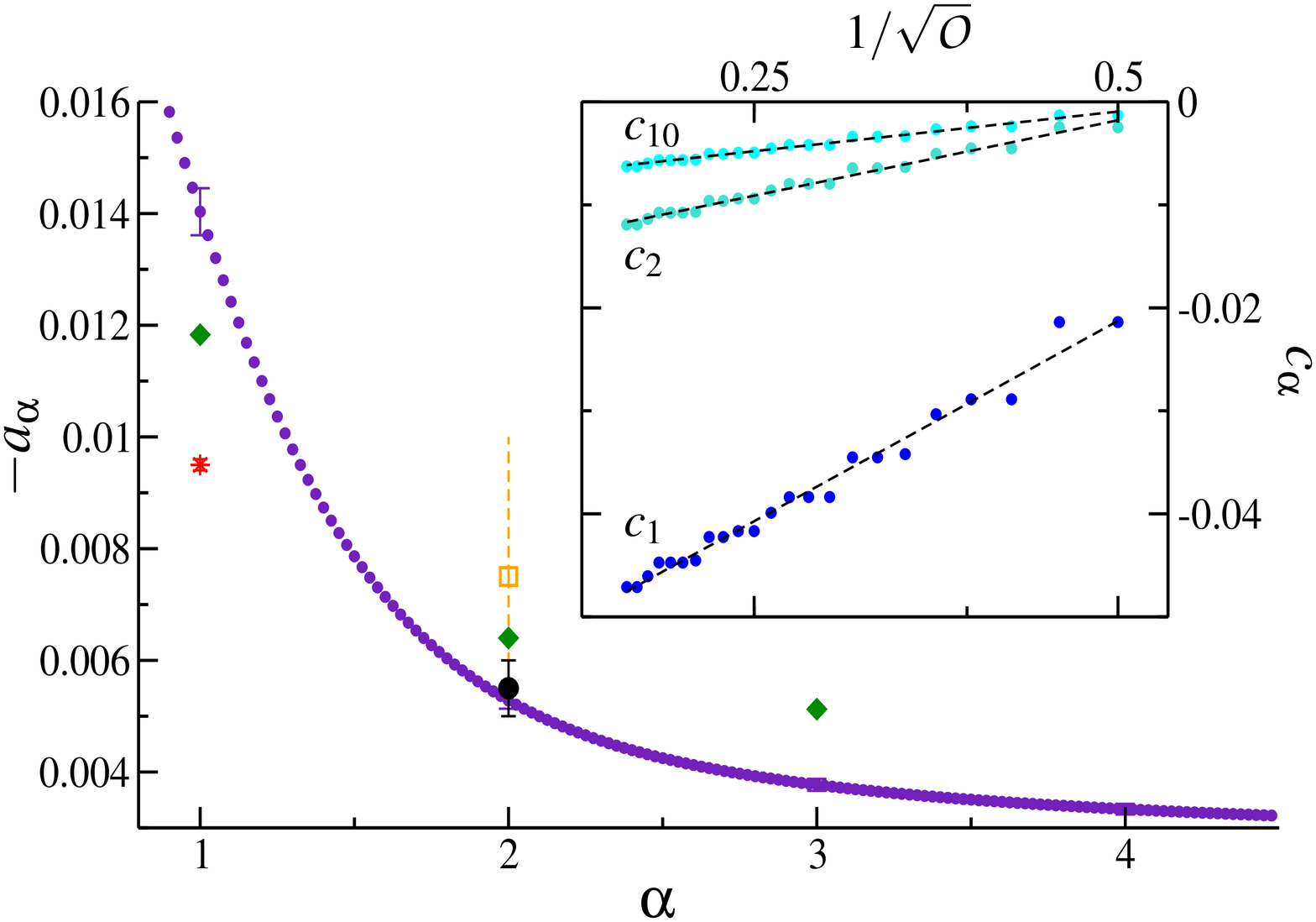}
\caption[]{The coefficient $-a_{\alpha}$ from fitting $c_{\alpha}$ to the function $a_{\alpha}\log\sqrt{\mathcal{O}}+b_{\alpha}$ (purple dots) plotted along with other numerical estimates of $-a_{\alpha}$ from free scalar field theory  \cite{logcorner} (green diamonds), tensor tree network  \cite{Luca} (red star), finite-T QMC \cite{Tommaso} (yellow square), and series expansion \cite{TFIM_series, footnote1} (black circle).
Standard error from the fit is shown for integer values of $\alpha$.
Inset: $c_{\alpha}$ for $\alpha=1,2,10$ along with fits plotted vs $1/\sqrt{\mathcal{O}}$ on a logarithmic scale.
\label{corner_crit}      
}
\end{figure}

The rest of the paper will focus on the quantum critical behavior.
Since high field NLCE is significantly more accurate for the thermodynamic limit, we restrict our study to $h \ge h_c$.
General arguments and our numerical study show that one can associate a length scale with
order $\mathcal{O}$ which goes as $\sqrt{\mathcal{O}}$.
In Fig.~\ref{corner_crit}, we show the behavior of the corner term $c_\alpha$ at $h=h_c$.
It is predicted to scale like $c_{\alpha} \propto a_{\alpha} \ln(L)$ \cite{logcorner}, 
with $a_{\alpha}$ being universal.
As shown in the inset of Fig.~\ref{corner_crit}, a plot of $c_{\alpha}$ versus $\ln(1/\sqrt{\mathcal{O}})$ can be 
extrapolated to the $\mathcal{O} \rightarrow \infty$ limit to obtain the universal term $a_\alpha$.
In the main plot of Fig.~\ref{corner_crit}, 
we show $a_{\alpha}$ as a function of Renyi index.  The values of this universal constant, as calculated by 
other methods, are also shown on the plot.  

The coefficient $a_2$ has been calculated for the 2D TFIM using several different numerical methods. In Ref.~\cite{TFIM_series}, series expansion found $a_2 = -0.0055(5)$ \cite{footnote1}. 
Finite-temperature QMC calculations on a single $36 \times 36$-size lattice report $-0.0075(25)$ \cite{Tommaso}.
Our current NLCE calculation gives 
$-0.0053$, with an uncertainty in the last digit -- a number consistent with the series expansion result.

This value of $a_2$ for the TFIM has been compared several times in the past literature to the value calculated analytically by Casini and Huerta 
for a free scalar field theory, $a_2 = -0.0064$ \cite{casiniComm}.  We note that, although this is numerically close to our value of $-0.0053$, in fact one should not expect correspondence since this is an interacting model that is not described by a free Gaussian fixed point.  Rather, one needs a calculation of the quantity in an interacting theory -- the lack of which we hope will motivate future field-theoretic calculations of $a_{\alpha}$ at the Wilson-Fisher fixed point. 

To our knowledge, the universal term $a_1$ has only been calculated once previously
using a tensor network variational ansatz called the ``tensor tree network'' (TTN), 
by Tagliacozzo and co-workers \cite{Luca}.  Their value for the universal 
coefficient is $a_1 = -0.0095(1)$.
The value from free scalar field theory is $-0.012$ \cite{logcorner}.
The value from the present NLCE work up to order $\mathcal{O}=26$ is 
$a_1 = -0.0140$, 
a number that is relatively close to the free field value.

Finally we note, based on the shape of the $a_{\alpha}$ versus $\alpha$ curve in relation to the
results from non-interacting theory in Fig.~\ref{corner_crit},
it is possible that $\alpha=2$ may not be the optimal Renyi index to make comparisons.
This is particularly important for methods such as QMC or series expansion that are restricted to integer $\alpha \ge 2$
due to reliance on the replica trick.  
In those cases, 
larger alpha values (e.g.~$S_3$) might be more suitable to distinguish the 
free scalar field theory from interacting ones.

Next, we examine the behavior of the line term $s_n$. Near the quantum critical point
the line entropy is expected to take the form \cite{Max},
\begin{equation}
S_2  = \eta L + S_0(L/\xi),
\end{equation}
where, the coefficient $\eta$ is non-universal but $S_0$ is a universal function.
In the limit $L/\xi \to 0$, it becomes a universal number $\gamma$ associated with
the critical point, while for $L/\xi \to \infty$, it takes the form $r_{\alpha} L/\xi$, with a universal
amplitude ratio $r_{\alpha}$.  
Like $c_{\alpha}$, this number may be used to distinguish different critical points by comparison between field theory, and
numerical calculation on model systems.  For the Gaussian fixed point, $r_{\alpha} \propto - \big(1 + \frac{1}{\alpha} \big)$ \cite{Cardy,Max}.
More interestingly, for the interacting field theory relevant to the TFIM, Ref.~\cite{Max} reports that
$r_{\alpha}$ changes sign near $\alpha=1$, a result consistent between $\epsilon$ and $1/N$ expansions at the Wilson-Fisher fixed point.  

\begin{figure}[t]
\includegraphics*[width=3.0in]{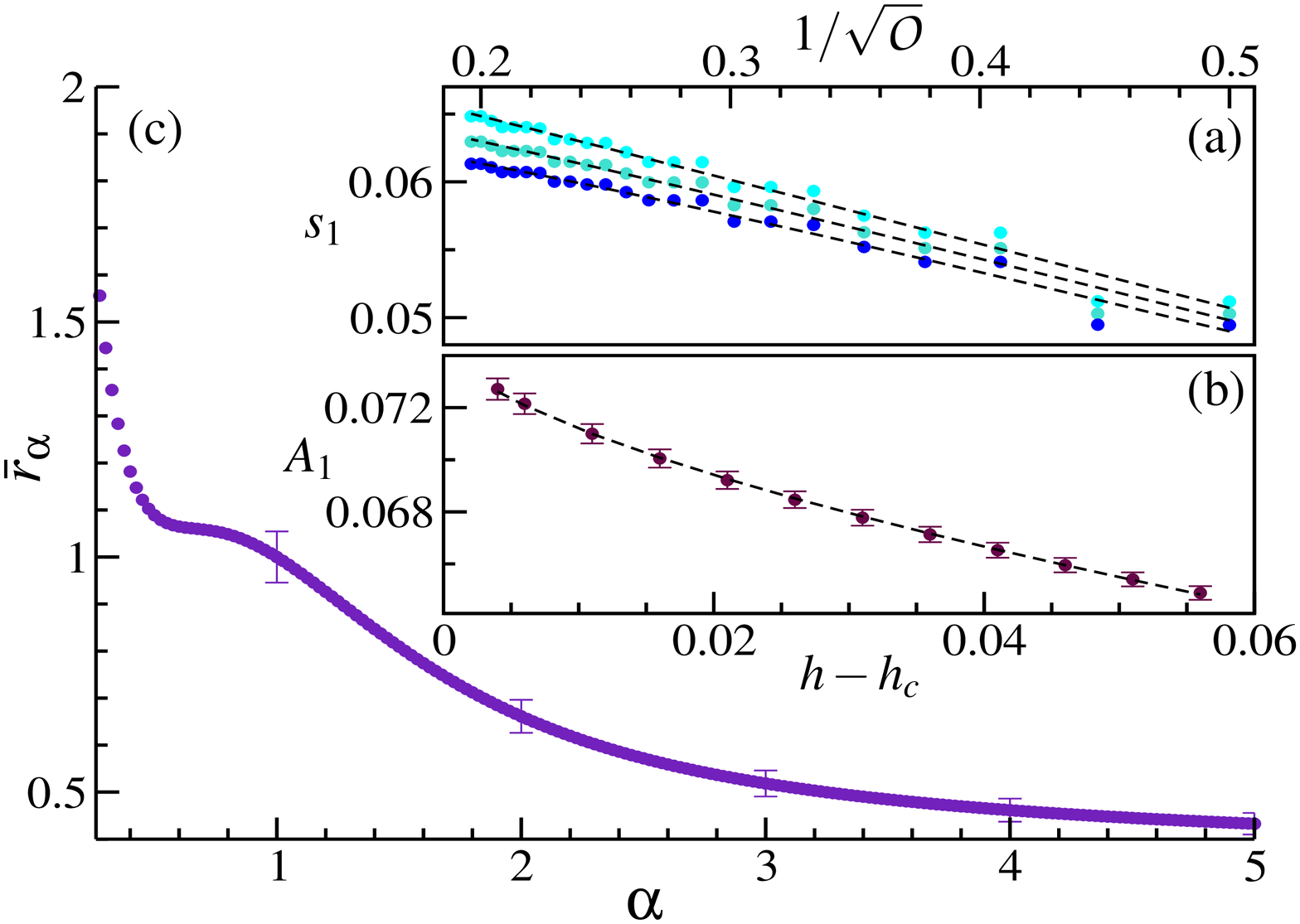}
\caption[]{
(a) Fits of the NLCE line term $s$ to Eq.~\eqref{fit1} for $\alpha=1$ from NLCE for $h=3.048$(top), $3.075$, and $3.1$(bottom).  
(b) The resulting linear term $A(h)$ from the fits in inset (a), fit to Eq.~\eqref{fit2}.  
(c) The normalized coefficient $\bar{r}_{\alpha}$ as a function of $\alpha$, with standard error shown for integer $\alpha$.
\label{line_crit}       
}
\end{figure}

To obtain $r_{\alpha}$, we first need to obtain $s_\alpha$ in the thermodynamic limit for $h$ near $h_c$.
We expect partial sums to order $\mathcal{O}$ to behave as $1/\sqrt{\mathcal{O}}$ if the correlation length of the system $\xi\sim |h-h_c|^{-\nu}$ is larger than $\sqrt{\mathcal{O}}$, but to saturate to the thermodynamic value when $\xi$ is less than $\sqrt{\mathcal{O}}$.
Hence, the NLCE data for $h$ near $h_c$ are fit for each $h$ and $\alpha$ to the function,
\beq
f(\mathcal{O},h) = A + \frac{B}{\sqrt{\mathcal{O}}}e^{-C\sqrt{\mathcal{O}}|h-h_c|^\nu},  \label{fit1}
\eeq
with $\nu=0.63$ \cite{IsingCritNu}. The constant $C$
was obtained from a fit to series expansion $s_2$ data \cite{TFIM_series}. 
Even if $C$ is allowed to vary in the NLCE fits, it does not change significantly as a function of $\alpha$.  
Thus it was held fixed at this value for all fits to Eq.~\eqref{fit1}, a few of which are shown in Fig.~\ref{line_crit}(a) for $s_1$ at three different values of $h$.

Figure \ref{line_crit}(b) shows an example of the resulting linear term, $A$ as a function of $h$ for $\alpha = 1$, from the fits to Eq.~\eqref{fit1}.  This term is then fit to,
\beq
A(h) = D + r_{\alpha}|h-h_c|^{\nu}, \label{fit2}
\eeq
for all values of $\alpha$.  This gives Fig.~\ref{line_crit}(c), showing the ratio $\bar{r}_{\alpha} = r_{\alpha}/r_1$
obtained after eliminating an unknown $\alpha$ independent prefactor in $r_\alpha$. 
It is apparent from this figure that 
there is no singularity or sign 
change in the universal amplitude $r_{\alpha}$ near the 
point $\alpha=1$. Our calculated $r_{\alpha}$ is negative for all $\alpha$, a result consistent with recent work of
Casini and Huerta on entanglement monotonicity \cite{circleEnt}, who argue that $r_1$ must always be negative \footnote{We thank Max Metlitski for pointing this out to us.}.

To independently confirm this result, we have used an alternative simpler fitting procedure 
to obtain $\bar{r}_\alpha$, using data only at the critical point $h_c$.
As discussed above, the quantity $\sqrt{\mathcal{O}}$ is a measure of the
correlation length $\xi$ explored (see Fig.~1) by NLCE. 
Hence, one can conjecture that the slope in the plot of $s_\alpha$ versus $\sqrt{\mathcal{O}}$
is proportional to $r_{\alpha}$ at $h_c$. 
Since there is an unknown proportionality constant between $\sqrt{\mathcal{O}}$ and $\xi$,  one can examine the ratio of the slope for a given Renyi index $\alpha$ with the slope at $\alpha=1$ to eliminate that constant.  The result of this procedure gives similar curves as in Fig.~\ref{line_crit}(c) for $\alpha>0.5$, 
with a slightly stronger divergence of $\bar{r}_{\alpha}$ for $\alpha<0.5$.  Again, although unusual non-monotonic behavior occurs around $\alpha \approx 1$, there is no singularity or sign change of $r_{\alpha}$ near the von Neumann point.

{\em Discussion --} 
In this paper we have developed a new NLCE procedure for calculating the bipartite Renyi entanglement entropy of
quantum lattice models using all $n\times m$ rectangular clusters at the interface of subsystems
$A$ and $B$.   The use of rectangular clusters, which can straight-forwardly be modified for other two-dimensional lattices,
overcomes the computational bottleneck arising from the exponentially-scaling subgraph isomorphism problem
that restricted previous NLCE methods to order ${\mathcal O} \le 16$.
With this innovation, our NLCE method is mainly constrained by memory usage of the exact diagonalization kernel.  The 
data in this study, obtained up to order 26 using less than 64GB memory, required only 1-2 CPU-months to complete.

Using the NLCE method, we have calculated individually the line and corner contributions of the Renyi
entanglement entropy of the transverse field Ising model for arbitrary Renyi index $\alpha$.
Extrapolating in the order of the calculation allows us to estimate universal critical
properties of this model.  
For entanglement across corners,
we conclusively demonstrate that a universal term in the $\alpha = 2$ entropy 
is distinct from the value calculated in a non-interacting field theory by Casini and Huerta \cite{logcorner}.
We hope this motivates future field-theory calculations at the interacting fixed point.
For entanglement across lines,
we searched for the striking sign change in the universal coefficient 
predicted by an interacting field theory near $\alpha = 1$ \cite{Max}, but find that no such sign change takes place,  
possibly implying a need for higher order $\epsilon$ or $1/N$ expansion terms in the field theory.
Rather, our data finds that this universal amplitude is always negative, consistent with recent theoretical arguments by Casini and Huerta 
based on entanglement monotonicity \cite{circleEnt}.

We have demonstrated that the accuracy of NLCE for the extraction of universal critical entanglement properties
rivals other numerical methods including quantum Monte Carlo, tensor tree networks, and series expansions.
In addition to the simplicity in graph counting introduced here, this success also stems from the fact that
NLCE can calculate Renyi entanglement entropies for arbitrary index $\alpha$,
and analytically separate contributions to the entanglement from line and corner geometries.

Perhaps the most appealing feature of the method is that this
accuracy results from a very simple numerical algorithm - less than a thousand lines of code are required to 
implement the rectangular-cluster NLCE into any existing Lanczos program.
Rectangular cluster geometries also give the exciting option of using the density matrix renormalization group \cite{White92} as a cluster solver, which could increase the accessible order of the NLCE to 100 or beyond.
Ultimately, its high accuracy, coupled with simplicity of implementation, lends hope that NLCE
will be widely adopted to the study of entanglement and other universal properties at quantum critical points 
in two and higher dimensions in the future.

{\em Acknowledgments --} We would like to thank M. Metlitski, T. Grover, A. J. Berlinsky, and I. Gonzalez for many
useful discussions.
This work was made possible by the computing facilities of Compute Canada via SHARCNET.  Support was provided 
by NSERC of Canada (ABK and RGM), the Ontario Ministry of Research and Innovation (ABK and RGM), and
by the NSF under grant numbers DMR-1004231 (RRPS) and 
PHY11-25915 (RGM).

\bibliography{Biblio}

\end{document}